\documentclass[11pt]{article}
\usepackage[dvips]{color}
\usepackage{mathtext}
\usepackage[koi8-r]{inputenc}
\usepackage{graphicx}
\usepackage[figuresright]{rotating}
\usepackage{amssymb,eucal}
\usepackage{cite}

\newcommand{\beq}{\begin{equation}}
\newcommand{\eeq}{\end{equation}}
\newcommand{\beqn}{\begin{eqnarray}}
\newcommand{\eeqn}{\end{eqnarray}}

\date{}
\begin{document}
\title{Geiger mode APD's for the underground cosmic ray experiment EMMA}

\author{L.~Bezrukov$^a$, K.~Butin$^a$, I.~Davitashvili$^a$, I.~Dzaparova$^a$, T.~Enqvist$^b$,\\ H.~Fynbo$^c$, L.~Golyshkin$^a$, Zh.~Guliev$^a$, L.~Inzhechik$^a$, 
A.~Izmaylov$^a$,\\
 J.~Joutsenvaara$^b$, M.~Khabibullin$^a$, A.~Khotjantsev$^a$, Yu.~Kudenko$^a$\footnote{Speaker.},\\
P.~Kuusiniemi$^b$, B.~Lubsandorzhiev$^a$, O.~Mineev$^a$, V.~Petkov$^a$,
R.~Poleshuk$^a$, \\
T.~R\"{a}ih\"{a}$^b$, J.~Sarkamo$^b$, B.~Shaibonov$^a$, A.~Shaykhiev$^a$, 
W.~Trzaska$^d$,\\
 G.~Volchenko$^a$, V.~Volchenko$^a$, A.~Yanin$^a$, N.~Yershov$^a$, D.~Zykov$^a$ \\
{}\\   
$^a${\it Institute for Nuclear Research RAS,  
117312 Moscow, Russia}\\  
$^b${\it CUPP/Pyh\"{a}salmi, University of Oulu, FIN-90014 Oulu, Finland} \\
$^c${\it Department of Physics and Astronomy, University of Aarhus,}\\ 
{\it DK-8000 {\AA}rhus, Denmark} \\
$^d${\it Department of Physics, University of Jyv\"{a}skyl\"{a},}\\
{\it FIN-40351 Jyv\"{a}skyl\"{a}, Finland} \\}
\maketitle

%
%
%
%
\begin{abstract}
{Multi-pixel photodiodes operating in a limited Geiger 
mode will be used for photoreadout of scintillator counters in underground 
cosmic ray experiment EMMA. Main parameters of photodiodes and 
the performance of EMMA scintillator counters are presented.}
\end{abstract}

%
%

\section{Introduction}
The goal of the EMMA experiment is to study the chemical composition 
of the primary cosmic rays around the ``knee'' region at the energy of 
about $3\times 10^{15}$  eV by measuring the multiplicity, lateral 
distribution
and arrival direction of the underground cosmic ray muons. The EMMA detector 
will be constructed in the Pyh\"{a}salmi mine, Finland~\cite{enqvist}. 
It
consists of drift chambers and plastic scintillator detectors and
has the total area of $\sim 135$ m$^2$ at the depth of 85 m (about
240 meters of water equivalent).

 In total, about 1600 scintillator counters, $122\times 122$ mm$^2$ and 3 cm
  thick, are arranged in $4\times 4$ arrays, each  of 16
  counters, which form individual   detectors of $50~{\rm cm}\times 50~{\rm cm}$ 
 in cross section.  Scintillator counters use wavelength shifting (WLS) 
 fiber readout. Photosensors, Geiger-mode multi-pixel avalanche photodiodes, 
 are mounted inside the counters. The set-up employing such photosensors is a unique 
  device designed for the  cosmic-ray induced underground muon detection.
 
\section{Parameters of MRS APD's}
\begin{figure}[h]
\centering\includegraphics[width=0.9\textwidth]{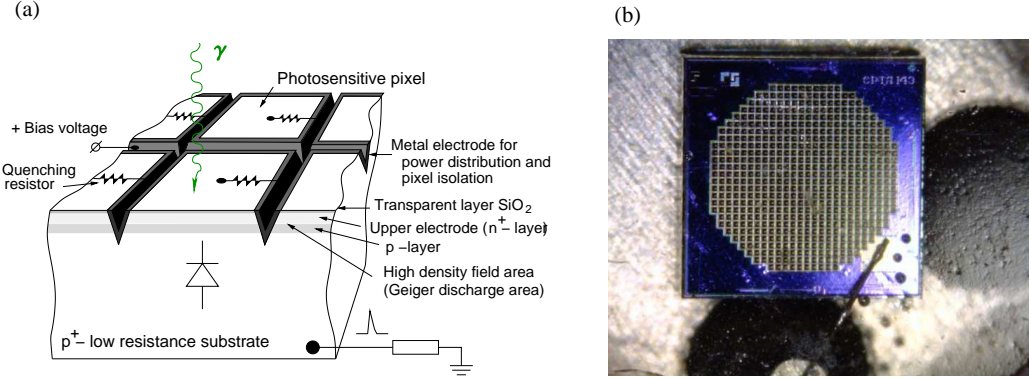}   
\caption{MRS APD: (a) schematic view of the structure; (b) photo of the
sensitive area.}
\label{fig:scheme_mrs}
\end{figure}

Detailed description of  multi-pixel avalanche photodiodes  with a  
metal$-$resistor$-$semiconductor  layer  structure  operating in a limited 
Geiger mode (hereafter referred to  as MRS APD's  or MRS photodiodes)  can be found in 
Refs.~\cite{dev-dop,dev2}.   Such a  photosensor  consists of many independent 
sensitive pixels produced on a common p$-$type silicon 
substrate.   A simplified topology of an MRS photodiode, invented and designed 
by the Center of Perspective Technologies and Apparatus (CPTA), Moscow, is 
shown in Fig.~\ref{fig:scheme_mrs}(a).
Each pixel operates as an independent Geiger  micro-counter with a gain of the 
same order as a vacuum photomultiplier. Pixels are separated by grooves filled 
with an optically non-transparent  material to suppress  an optical  
cross$-$talk. The  MRS photodiodes with  556 pixels of 
$35\times35~\mu {\rm m}^2$ each and total  sensitive area of 1.1~mm$^2$ (see 
Fig.~\ref{fig:scheme_mrs}(a))
were manufactured by CPTA for the EMMA experiment. Main parameters of these 
MRS APD's are reviewed below.

{\it Bias voltage}. The bias voltage, $V_{bias}$, applied on a photodiode creates  a very 
high electric field in a very thin  layer ($\sim$1~$\mu$m p-n junction).  
The bias voltage is set to be above the breakdown level, $V_{bd}$, so that 
the electric 
field can sustain 
the carrier's avalanche; however there is no current through the depletion layer 
until the first carrier  is generated. Value of overvoltage 
$\Delta V = V_{bias} - V_{bd}$ determines the  gain, 
photon detection  efficiency (PDE),  intrinsic noise and cross--talk. Typical 
bias voltage of MRS APD's for EMMA detectors  is around 30-32~V. We apply the 
overvoltage of 4~V  in operating mode for tests. EMMA electronics will
automatically  tune the bias voltages to have  the background signals  
in the detector roughly at the same rate for all  counters.

{\it Photon detection efficiency}. The PDE was measured at room temperature 
with a spectrophotometer and a  calibrated PMT as described 
in~\cite{akhrameev,musienko}.  Our result for PDE is about 25\%  for green 
light region (WLS fiber emission spectrum)  at the operating bias voltage. 
The PDE values of MRS APD's in green light region is close to that of  
Hamamatsu MPPC's manufactured for the  T2K experiment~\cite{yokoyama,retiere}. 
We have come to this conclusion after comparative  measurements between 
MRS APD's and MPPC's with a reference light source.

\begin{figure}[hbt]
\centering\includegraphics[width=12cm,angle=0]{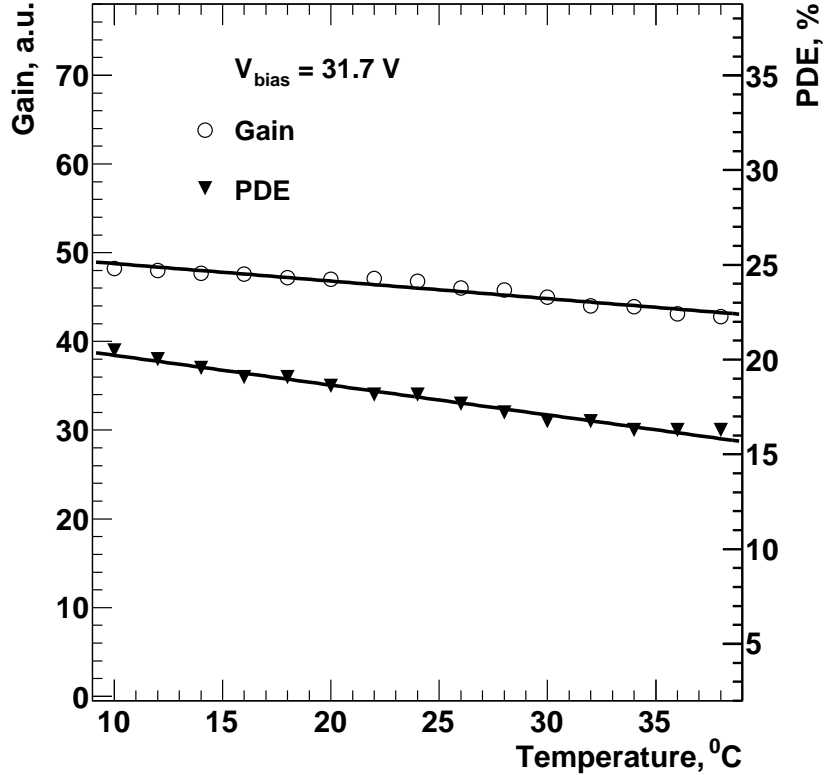}
\caption{The photon detection efficiency for green light 
($\lambda \sim 515$ nm)  and the gain of an MRS APD as a function of   
the temperature.}
\label{fig:pde_gain_temp} 
\end{figure}

\begin{figure}[hbt]
\centering\includegraphics[width=10cm,angle=0]{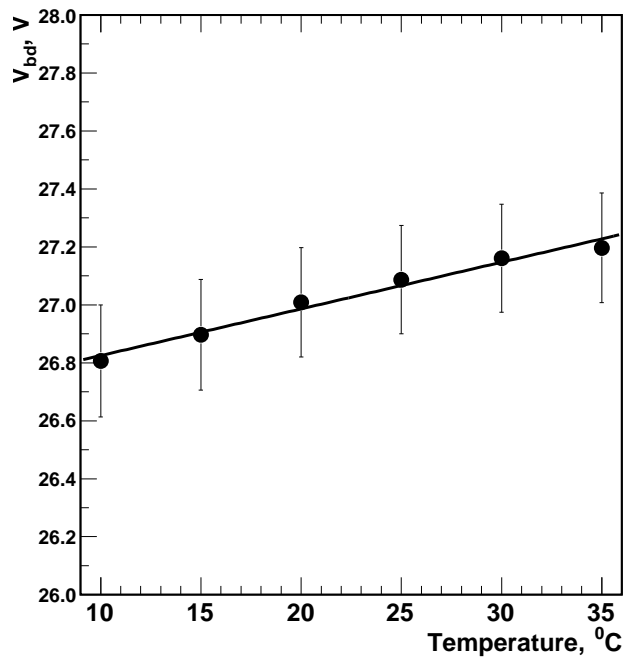}
\caption{Temperature dependence of the breakdown voltage. The solid line 
shows  the best fit with a gradient of 16~mV/$^{\circ}$C.}
\label{fig:vb_temp} 
\end{figure}

{\it Temperature dependence.} Since the breakdown voltage  depends on 
temperature the main parameters are sensitive to  ambient temperature. 
Parameters of an MRS photodiode (PDE and gain)  were measured inside a 
thermostatic box in a wide temperature 
interval from 10 to 38~$^{\circ}$C.  A WLS Y11 
fiber excited by a blue LED was used as a light source. The stability 
of LED light  
was measured to be better than 0.1~\%/$^{\circ}$C. The results of the
measurements are shown in Fig.~\ref{fig:pde_gain_temp}. 
The MRS APD gain   decreases with temperature as   -0.4~\%/$^{\circ}$C, 
while the PDE varies with temperature  as -0.9~\%/$^{\circ}$C. The  
weak sensitivity of MRS APD parameters to temperature 
can be explained by two factors: a weak temperature dependence of 
the breakdown voltage and a relatively high $\Delta V$.  
$V_{bd}$ was 
measured in the range  from 10  to 35~$^{\circ}$C. As  
seen from Fig.~\ref{fig:vb_temp}, 
breakdown voltage shows  a weak  temperature dependence with 
a gradient of about 16~mV/$^{\circ}$C.

{\it Pulse shape}. The pulse shape   of  this type of photosensor is 
determined by a quenching resistor  which is  about 20~M$\Omega$.  
This value, which affects the pulse shape, was not well controlled during the 
production of wafers.
The pulse shape of some MRS APD's  has a long tail up to a few $\mu$s, as 
discussed in Ref.~\cite{pd07}. To overcome such a problem, all MRS APD's were 
separated into 4 groups according to 
the ratio of the fast and slow (tail) components.  
In addition, a large quenching resistor causes  the long recovery time of 
about 1 $\mu$s. 
Since the  expected counting rate   in the  mine, including  background from 
the surrounding 
rock, will be $\leq 100$ Hz per a single counter~\cite{volchenko}, the
recovery time is not a
problem for the EMMA experiment.  On the other hand, the 
long recovery time  allows us to avoid after-pulses in contrast to  
MPPC's which were designed to have a fast recovery time~\cite{retiere_recovery}.

In total, about 2200 MRS photodiodes fabricated from a few different wafers 
were tested. About 10~\% of delivered devices were rejected as defective 
(no signal) during 
preliminary tests and returned back  to the manufacturer. No more dead 
MRS APD's were found during tests after acceptance. 
In order  to meet the requirements of the experiment,  the operating bias 
voltages  were 
set so that the dark rate of each device 
was close to a reference   value of   about 1~MHz at  
22~$^{\circ}$C   at a  
discriminator threshold of 0.5~p.e. The breakdown voltages for these 
MRS APD's   were found to be in a wide range $25 - 31$ V that  determines 
the range  of $29 - 35$~V for the applied bias voltages    
to provide the PDE value of $\geq 20$\% for green light.  
  
\section{EMMA counters}
\begin{figure}[hbt]
\centering\includegraphics[width=0.9\textwidth]{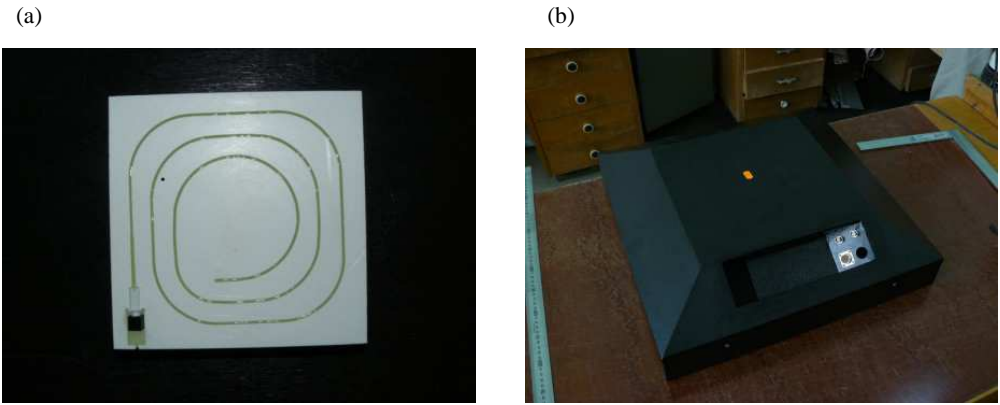}   
\caption{Photographs: (a) an individual scintillator counter with glued Y11 
fiber and embedded optical 
connector;  (b) an EMMA  module comprised of 16 counters in a $4\times 4$ assembly. }
\label{fig:counter_module}
\end{figure}

An EMMA scintillator counter is shown in Fig.~\ref{fig:counter_module}(a). 
Polystyrene scintillator tiles ($122\times 122\times 30$  ${\rm mm}^3$)  
have 
been manufactured  at IHEP, Protvino, Russia. Scintillating 
light is collected by a  single Y11 (Kuraray) WLS fiber of 1 mm diameter 
glued into a spiral
groove  of 3.2 mm depth with the BC600 Bicron glue.  The groove was  carved by 
a Woodpecker  engraving-milling machine  which speed and cutting tools were 
optimized 
to make  a clean groove to provide the good transmission of the scintillating 
light 
through the cut surfaces.   
This technique was developed for scintillator counters for muon 
detection~\cite{smrd} in the near neutrino  detector of  the long baseline 
neutrino  experiment T2K~\cite{nd280}.  The fiber is viewed from 
one end by an MRS photodiode, the other end is covered by an aluminized mylar 
reflector.  
This readout configuration allows us  to collect scintillation light uniformly over 
the entire scintillator.  
Outer surfaces of scintillators were  etched by a chemical agent that 
resulted in the formation of a micropore 
deposit over the plastic surface. The thickness of the 
deposit ($30 - 100$ $\mu$m), which works as a diffuse reflector,  depends on 
the etching  time.   Details can be found 
in Ref.~\cite{extrusion}. Scintillators with glued WLS fibers were wrapped 
into an additional reflector layer of 0.1 mm thick Tyvek paper that 
increases a light yield by about 15\%. 
One module is comprised of 16 scintillator
counters ($4\times 4$ assembly) which are packed  into a steel 
box of $0.5\times 0.5$ ${\rm m}^2$, as shown in 
Fig.~\ref{fig:counter_module}(b). 

\begin{figure}[hbt]
\centering\includegraphics[width=0.9\textwidth]{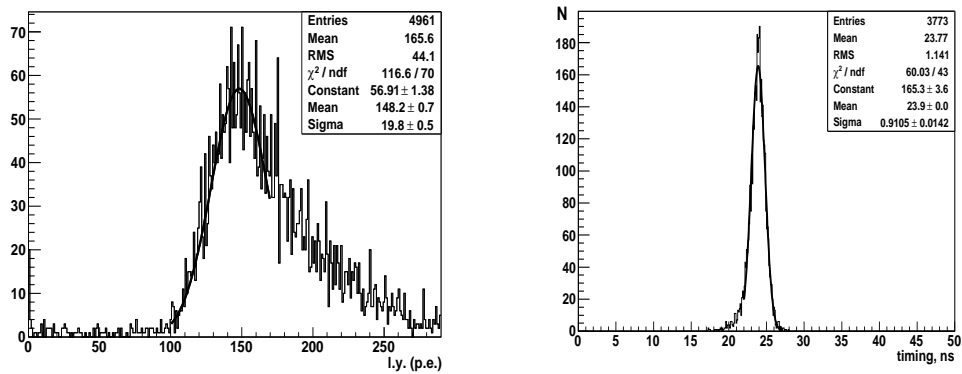}   
\caption{Left: the cosmic muon spectra measured by a typical counter. The light yield
of 150 p.e. corresponds to the deposited energy of about 6 MeV. Right: time 
resolution for cosmic muons.}
\label{fig:muon_spectra}
\end{figure}

The performance of EMMA counters was tested with cosmic 
ray muons   using a small 2$\times$2 cm$^2$  
trigger counter 
placed at the  center of a tested counter. A typical muon spectrum is shown in 
Fig.~\ref{fig:muon_spectra}. 
The light yield of about 150 photoelectrons was obtained for a 6 MeV energy 
deposited by cosmic muons in the 3 cm thick counter.
The obtained time resolution ($\sigma$) is about 0.9 ns 
(see Fig.~\ref{fig:muon_spectra}) and  mainly 
determined by the slow decay time (9-12 ns) of the Y11 fiber. 

\begin{figure}[htb]
\centering\includegraphics[width=0.8\textwidth]{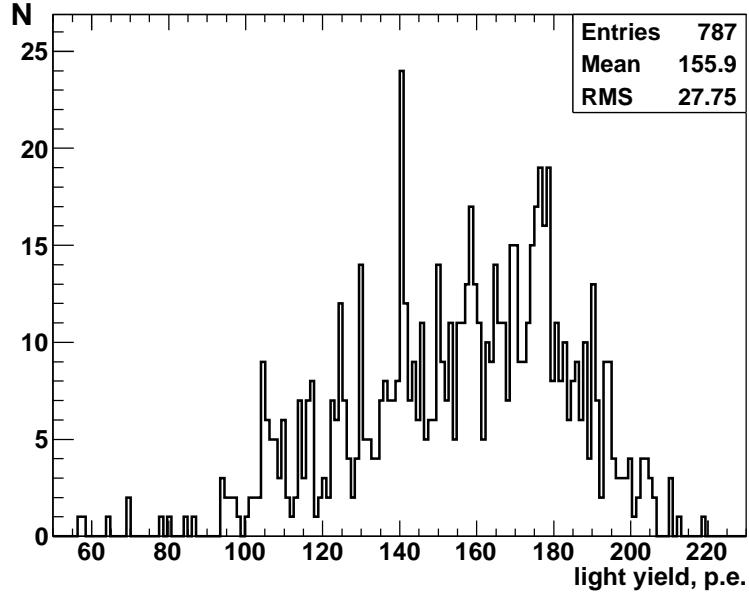}   
\caption{The light yield distribution of manufactured  EMMA counters. }
\label{fig:ly_distr}
\end{figure}

EMMA counters must provide a high light yield to separate cosmic muon signals  
from photon background of the rock walls  ($E_{\gamma} =2.6$  MeV). The light
yield of more than 
100 p.e. per cosmic muon allows us to  suppress the photon background and keep a 
high efficiency ($>95$\%) for cosmic muons. The light yield distribution for 
about 800  counters is shown in  Fig.~\ref{fig:ly_distr}. 
Almost all manufactured counters have the light yield 
of  more  
than  100 p.e. and satisfy the requirements of the experiment.

\section{Conclusion}
The MRS APD's manufactured for the EMMA experiment have  high PDE values
and show a low temperature dependence of the main parameters.   
The scintillator counters with WLS fiber readout and MRS APD's  provide a 
high light yield  and a good time resolution. The production of all counters 
will be finished in December 2009. Installation of all modules and 
commissioning of the 
detector is expected in the first half of 2010. 

 It is a pleasure for authors to thank the INR technicians for their excellent 
work on the preparation and  assembly of the detectors. The work of the 
authors from the University of Oulu is funded by
the European Union Regional Development Fund and is also supported
by the Academy of Finland (Project 108991).


\begin{thebibliography}{99}
 \bibitem{enqvist}T.~Enqvist {\it et.al.,} Nucl. Phys. B (Proc. Suppl.) 
{\bf 165} (2007) 349; 
T.~Enqvist {\it et.al.,} Nucl. Phys. B (Proc. Suppl.) 
{\bf 175-176} (2008) 307.
\bibitem{dev-dop}V.~Golovin {\it et.al.,} Patent for invention in Russia, 
No. 1644708, 1989;
A.G.~Gasanov {\it et al.,} Lett. J. Techn. Phys. {\bf 16} (1990) 
14  (in Russian). 
\bibitem{dev2}G.~Bondarenko {\it et al.,} Nucl. Instr.  Meth., 
{\bf A442} (2000) 187.
\bibitem{akhrameev}E.V.~Akhrameev {\it et al.,} arXiv:0901.4675 
[physics.ins-det].
\bibitem{musienko}Yu.V. Musienko {\it et al.,} Instrum. Exp. Tech. 
{\bf 51} (2008) 101.
\bibitem{yokoyama}M.~Yokoyama {\it et al.,} arXiV:0807.3145 [physics.ins-det].
\bibitem{retiere}F.~Retiere, \emph{Using MPPC's for T2K near
detector}, talk at this Workshop.
\bibitem{pd07}Yu.Kudenko et al., PoS PD07:016, 2007.
\bibitem{volchenko}V.I.~Volchenko {\it et al.,} 
arXiV:0810.2414 [physics.ins-det].
\bibitem{retiere_recovery} Y.~Du and F.~Retiere, Nucl. Instr. Meth. 
{\bf A596} (2008) 396.
\bibitem{smrd}O.~Mineev {\it et al.,} Nucl. Instr. Meth. {\bf A577} (2007) 540;
A.Izmaylov {\it et al.,} arXiv:0904.4545.
\bibitem{nd280}``T2K ND280 Conceptual Design Report'',  T2K Internal Document;
D.~Karlen, Nucl. Phys. B (Proc. Suppl.) {\bf 159} (2006) 91;
Yu.G.~Kudenko, Nucl. Instrum. Meth. {\bf A598} (2009) 289.
\bibitem{extrusion}Yu.G.~Kudenko {\it et al.,}  Nucl. Instr. Meth. {\bf A469} (2001) 340; 
O.~Mineev {\it et al.}, Nucl. Instr. Meth. {\bf A494} (2002) 362; 
N.~Yershov {\it et al.}, Nucl. Instr. Meth. {\bf A543} (2005) 454. 
 
 
\end{thebibliography}
\end{document}